\documentclass[aps,pra,preprint,showpacs,superscriptaddress]{revtex4-1}

\usepackage{tikz}
\usepackage{pgfplots}
\usetikzlibrary{plotmarks}
\usetikzlibrary{arrows,decorations.pathmorphing,backgrounds,positioning,fit,petri}
\usetikzlibrary{calc} 

\usepackage{hyperref}
\hypersetup{
	bookmarksopen=true,%
	bookmarksnumbered=true,%
	colorlinks=true,%
    	linkcolor=blue,
    	citecolor=blue,
    	filecolor=blue,
    	urlcolor=blue,
	pdfstartview=FitH,%
	pdfnewwindow=true
}

\usepackage{graphicx}

\usepackage{mathtools} 
\mathtoolsset{showonlyrefs}

\usepackage{txfonts}

\usepackage{microtype}

\newcommand{\bra}[1]{\langle#1\rvert}
\newcommand{\ket}[1]{\lvert#1\rangle}
\newcommand{\qprod}[2]{ \langle #1 | #2 \rangle }
\newcommand{\braopket}[3]{\langle #1 | #2 | #3\rangle}

\newcommand{\soko}{s^{(1)}_{k_1}}
\newcommand{\sokt}{s^{(1)}_{k_2}}
\newcommand{\stko}{s^{(2)}_{k_1}}
\newcommand{\stkt}{s^{(2)}_{k_2}}

\makeatletter
\newcommand{\vast}{\bBigg@{2}}
\newcommand{\Vast}{\bBigg@{4}}
\makeatother

\newcommand{\me}{\mathrm{e}}
\newcommand{\mi}{\mathrm{i}}

\newcommand{\dif}{\mathrm{d}}

\newcommand{\abs}[1]{\lvert#1\rvert}

\begin{document}
\title{Few photon transport in a waveguide coupled to a pair of collocated two-level atoms}
\author{Eden Rephaeli}
\email{edenr@stanford.edu}
\affiliation{Department of Applied Physics, Stanford University, Stanford, CA 94305}
\author{ \c{S}\"ukr\"u Ekin Kocaba\c{s}}
\author{Shanhui Fan}
\email{shanhui@stanford.edu}
\affiliation{Department of Electrical Engineering, Stanford University, Stanford, CA 94305}

\date{\today}

\begin{abstract}
We calculate the one- and two-photon scattering matrices of a pair of collocated non-identical two-level atoms coupled to a waveguide. We show that by proper choice of a two-photon input, the background fluorescence by the atoms may be completely quenched, as a result of quantum interference, and that when the atoms' detuning is smaller than their linewidths, extremely narrow fluorescence features emerge.  Furthermore, the system emits a two-photon bound state which can display spatial oscillations/quantum beats, and can be tuned from bunched to anti-bunched statistics as the total photon energy is varied.
\end{abstract}

\maketitle


\pdfbookmark[1]{Introduction}{Introduction}
\section{Introduction} There has been substantial recent interest in the study of photon-atom interactions, where microwave \cite{Wallraff2004,Shen2005,Shen2007,Astafiev2010,Gambetta2011} and optical \cite{Aoki2006,Akimov2007,Shen2007a,Zhou2008,Rephaeli2010,Longo2010,Shi2011,Roy2010,Liao2010,Zheng2010,Gonzalez-Tudela2011,Kolchin2011} photons are confined to a single-mode waveguide. From a practical point of view, quantum states of light are important carriers of information in quantum information and quantum computing systems.  The use of waveguides to connect qubits can enable entanglement transfer \cite{Cirac1997,Gonzalez-Tudela2011}, and is important for integration. From a more basic point of view, the one-dimensional nature of photon states in a single-mode waveguide leads to a number of novel physics effects in photon-atom interactions, as well as device possibilities.  For example, confinement of photons to one dimension enables the complete reflection of a single photon from a two-level atom \cite{Shen2005}, and the full inversion of an atom with a single-photon pulse \cite{Rephaeli2010}. Furthermore, it has been shown that when two photons scatter off a two-level system, a photon-photon bound state and an associated background fluorescence emerge \cite{Shen2007}. Recently, logic operations at the single photon level through the use of a three-level system in a waveguide have been investigated \cite{Chang2007}. 

\begin{figure}[tb]
\begin{center}
\includegraphics[scale=0.6]{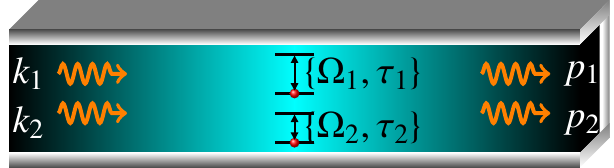}
\caption{(Color online) Schematic representation of a pair of nonidentical two-level atoms in a waveguide geometry. $k_{1,2}$  and $p_{1,2}$ denote the incoming and scattered photons, respectively.  $\Omega_{1,2}$ and $\tau_{1,2}$ are the transition frequencies and decay times of the two atoms, respectively.}
\label{figGeometry}
\end{center}
\end{figure}

In this work, we consider a waveguide coupled to two non-identical quantum two-level systems, as shown in Fig. \ref{figGeometry}, and study few-photon transport. By two-level atoms, we are primarily concerned with on-chip atomlike objects, such as superconducting qubits \cite{Wallraff2004} or quantum dots \cite{Akimov2007}.  These objects can be described by the same two-level Hamiltonian as a real atom, hence our nomenclature.  However, in contrast to real atoms, one of the distinct properties of these atomlike objects is their tunability \cite{Haft2002,Hogele2004,Faraon2007,Gambetta2011}.  Our predicted effects exploit the tunability of these atomlike objects.
 
Here we show that the two-atom system can be solved exactly in the two-photon Hilbert space, using input-output formalism \cite{Gardiner1985} adapted for the calculations of few-photon Fock-state transport \cite{Fan2010}. The results point to a rich set of physics, some of which may be important to device applications.  Previously, this system was studied at the one-photon level \cite{Kim2010,Shen2007a}, and was shown to exhibit a single-photon transmission spectrum that is the analogous to electromagnetically induced transparency (EIT) \cite{Shen2007a}. Quantum entanglement and modification of spectral features via a resonant laser \cite{Das2008}, as well as spatial modulation of spontaneous emission decay \cite{Zanthier2006} were both studied in two \textit{identical} two-level atoms in free space. Multiphoton scattering in a waveguide was studied only in multiple \textit{identical} atoms \cite{Yudson2008}.    Here we show that allowing for non-identical atoms enables fluorescence linewidth narrowing and quenching, as well as new capabilities to design and control the properties of photon-photon bound states.  These two-photon bound states can exhibit bunching or anti-bunching statistics.  The two photons forming the bound states can have very different frequencies.  Moreover, the properties of the bound state, including its spatial extent, are strongly dependent on the resonant frequencies of the atoms.  None of these characteristics have been observed in waveguide-atom systems consisting of either a single two-level atom or a single three-level atom. 

Two-photon bound states represent a composite particle of photons, and are of substantial interest in quantum lithography and imaging \cite{Dangelo2001}.  Our results show that in the two-atom system there is enhanced capability for generation and control of such a composite quantum object.  Also, one typically expects two-photon bound states to arise from effective photon-photon attraction, and thus one typically expects the two-photon bound state to exhibit a bunching behavior.  Indeed, only bunched two-photon bound states have been seen in all waveguide-atom \cite{Shen2007, Roy2011}, and waveguide-nonlinear-cavity \cite{Liao2010} systems previously considered.  In this context, our result, showing an anti-bunched two-photon state, is counter-intuitive, and points to the substantial richness in physics of photon-photon interaction in the two-atom system that is qualitatively different from all previously considered systems.

\pdfbookmark[2]{System Hamiltonian and equations of motion}{System Hamiltonian and equations of motion}
\section{System Hamiltonian and equations of motion}
\label{System hamiltonian and equations of motion}
For the system shown in Fig. \ref{figGeometry}, the waveguide supports both left and right propagating photon modes, though the photon-atom interaction is entirely contained in the even subspace \cite{Shen2007}.  This subspace features a chiral photonic band interacting with two atoms, which in the rotating-wave approximation is described by the Hamiltonian: ($\hbar=1$; the waveguide group velocity is set to $v_g=1$)
\begin{align}
\hat{H}=&\int \dif k\ k\ a^{\dagger}_k a_k+V_1\int \dif k \left[ a^{\dagger}_k\sigma^{(1)}_-+\sigma^{(1)}_+a_k \right]+\frac{1}{2}\Omega_1\sigma^{(1)}_z
\\&+\frac{1}{2}\Omega_2\sigma^{(2)}_z +V_2\int \dif k \left[ a^{\dagger}_k\sigma^{(2)}_-+\sigma^{(2)}_+a_k \right] \tag{1}\label{E:Htotal}
\end{align}
where $\Omega_1$ and $\Omega_2$ are the atoms' transition frequencies; $ V_{1,2}$ are their respective coupling strengths to the waveguide field, and are related to the spontaneous decay rate of each atom by $\frac{1}{\tau_{1,2}}=\pi V^2_{1,2}$.  $a_k \ (a^{\dag}_k) $ destroys (creates) a waveguide photon with energy $k$, and satisfies $ [a_{k},a^{\dag}_{k'}]=\delta(k-k')$.  $\sigma_\mp^{(1,2)}$ are the lowering and raising operators for each atom.  Once the S-matrix of the chiral mode is determined, the transport properties of the system in Fig.\ref{figGeometry} may be obtained using standard techniques \cite{Shen2007}.

To solve the few-photon Fock-state transport properties for the Hamiltonian in Eq. \eqref{E:Htotal}, following Ref. \cite{Fan2010} we define the input, $a_{\text{in}}(t)=(2\pi)^{-1/2}\int \dif k\me^{-\mi k(t-t_0)}a_k(t_0)$, and output, $a _{\text{out}}(t)=(2\pi)^{-1/2}\int \dif k\me^{-\mi k(t-t_1)}a_k(t_1)$, operators.  Here, $t_0$ and $t_1$ refer to times long before $(t_0\rightarrow-\infty)$ and long after $(t_1\rightarrow+\infty)$ the photons interact with the atoms. Following the procedure in \cite{Gardiner1985,Fan2010}, we arrive at the input-output fomalism equations
\begin{align}
&a_{\text{out}}(t)=a _{\text{in}}(t)-\mi\sqrt{\frac{2}{\tau_1}}\sigma^{(1)}_-(t)-\mi\sqrt{\frac{2}{\tau_2}}\sigma^{(2)}_-(t), \label{E:aoutain} 
\\&\frac{\dif\sigma^{(1)}_-(t)}{\dif t}={-\mi\left(\Omega_1-\mi\frac{1}{\tau_1}\right)}\sigma^{(1)}_-(t)+\mi\sqrt{\frac{2}{\tau_1}}\sigma^{(1)}_z(t) a _{\text{in}}(t)+\frac{\sigma^{(1)}_z(t)\sigma^{(2)}_-(t)}{\sqrt{\tau_1\tau_2}}, \label{E:sig1b}
\\&\frac{\dif\sigma^{(2)}_-(t)}{\dif t}={-\mi\left(\Omega_2-\mi\frac{1}{\tau_2}\right)}\sigma^{(2)}_-(t)+\mi\sqrt{\frac{2}{\tau_2}}\sigma^{(2)}_z(t) a _{\text{in}}(t) +\frac{\sigma^{(2)}_z(t)\sigma^{(1)}_-(t)}{\sqrt{\tau_1\tau_2}}. \label{E:sig2b}
\end{align}
We note that the inclusion of a dipole-dipole interaction term in Eq. \eqref{E:Htotal} of the form $g\left(\sigma^{(1)}_+\sigma^{(2)}_-+\sigma^{(2)}_+\sigma^{(1)}_-\right)$, where $g$ is the coupling rate,  may be accounted for by making the replacement $(\tau_1\tau_2)^{-1/2}\mapsto (\tau_1\tau_2)^{-1/2}+\mi g$ in the last term in Eqs. \eqref{E:sig1b} and \eqref{E:sig2b}.  The solution in this case would carry on in a similar fashion.  Here, we exclude the dipole-dipole term.

Below, we will solve Eqs. \eqref{E:sig1b} and \eqref{E:sig2b} to obtain the single-photon scattering amplitudes $\qprod{p^-}{k^+}\equiv\braopket{0}{a_\text{out}(p)a^{\dag}_\text{in}(k)}{0}$, and the two-photon scattering amplitude $\qprod{p_1 p^-_2}{k_1 k^+_2}=\braopket{0}{a_\text{out}(p_1)a_\text{out}(p_2)a^{\dag}_\text{in}(k_2)a^{\dag}_\text{in}(k_1)}{0}$.  Here the $k$'s and $p$'s are incident and outgoing free-photon energy, respectively.
\pdfbookmark[3]{One-Photon Scattering Matrix}{One-Photon Scattering Matrix}{One-Photon Scattering Matrix}{One-Photon Scattering Matrix}
\section{One-Photon Scattering Matrix} Fourier transforming of Eq. \eqref{E:aoutain} leads to the single-photon scattering amplitude
\begin{align}
&\bra{p^-}k^+\rangle=\bra{0}a_{\text{in}}(p)\ket{k^+}-\mi\sum_{n=1,2}\sqrt{\frac{2}{\tau_n}}\bra{0}\sigma^{(n)}_-(p)\ket{k^+} \label{E:SoneIn}
\end{align}
In order to calculate $\braopket{0}{\sigma^{(1,2)}_-(p)}{k^{+}}$, we use Eqs. \eqref{E:sig1b} and \eqref{E:sig2b} and solve for $\bra{0}\sigma^{(1,2)}_-(t)\ket{k^+}$. The solution results in $\bra{0}\sigma^{(1,2)}_-(p)\ket{k^+} = s^{(1,2)}_k \delta(k-p)$, where
\begin{align}
&s^{(1,2)}_k = \sqrt{\frac{2}{\tau_1}}\frac{\left(k-\Omega_{2,1}\right)}{\left(k-\Omega_1+\mi\frac{1}{\tau_1}\right)\left(k-\Omega_2+\mi\frac{1}{\tau_2}\right)+\frac{1}{\tau_1\tau_2}}. \label{E:s1k}
\end{align}
are the excitation amplitudes of the two atoms.
We plug Eq. \eqref{E:s1k} into Eq. \eqref{E:SoneIn} to obtain the one-photon scattering matrix $\qprod{p^-}{k^+}=t_k\delta(k-p)$ where
\begin{align}
&t_k=\frac{\left(k-\Omega_1-\mi\frac{1}{\tau_1}\right)\left(k-\Omega_2-\mi\frac{1}{\tau_2}\right)+\frac{1}{\tau_1\tau_2}}{\left(k-\Omega_1+\mi\frac{1}{\tau_1}\right)\left(k-\Omega_2+\mi\frac{1}{\tau_2}\right)+\frac{1}{\tau_1\tau_2}}.
\label{E:Skp}
\end{align}

\pdfbookmark[4]{Two-Photon Scattering Matrix}{Two-Photon Scattering Matrix}
\section{Two-Photon Scattering Matrix} 
The two-photon scattering matrix can be written as
\begin{align}
&\braopket{0}{a _\text{out}(p_1) a_\text{out}(p_2) a^\dag_\text{in}(k_2) a^\dag_\text{in}(k_1)}{0}
=t_{p_1}\braopket{p^{+}_1}{a _{\text{in}}(p_2)-\mi\sqrt{\frac{2}{\tau_1}}\sigma^{(1)}_-(p_2)-\mi\sqrt{\frac{2}{\tau_2}}\sigma^{(2)}_-(p_2)}{k_1 k_2^{+}} \label{E:Stwo}
\end{align}
where we have used Eq. \eqref{E:SoneIn}. We are then tasked with calculating the matrix elements $\braopket{p^+_1}{\sigma^{(1,2)}_-(p_2)}{k_1 k^+_2}$.  For this purpose we again use Eqs. \eqref{E:sig1b}--\eqref{E:sig2b} to get a coupled differential equation for $\braopket{p^+_1}{\sigma^{(1,2)}_-(t)}{k_1 k^+_2}$ with inhomogenous terms of the type $\braopket{p^+_1}{\sigma^{(1,2)}_z(t) a _\text{in}(t)}{k_1 k^+_2}$ and $\braopket{p^+_1}{\sigma^{(1)}_z(t)\sigma^{(2)}_-(t)}{k_1 k^{+}_2}$.
\begin{figure}
\includegraphics[scale=0.5]{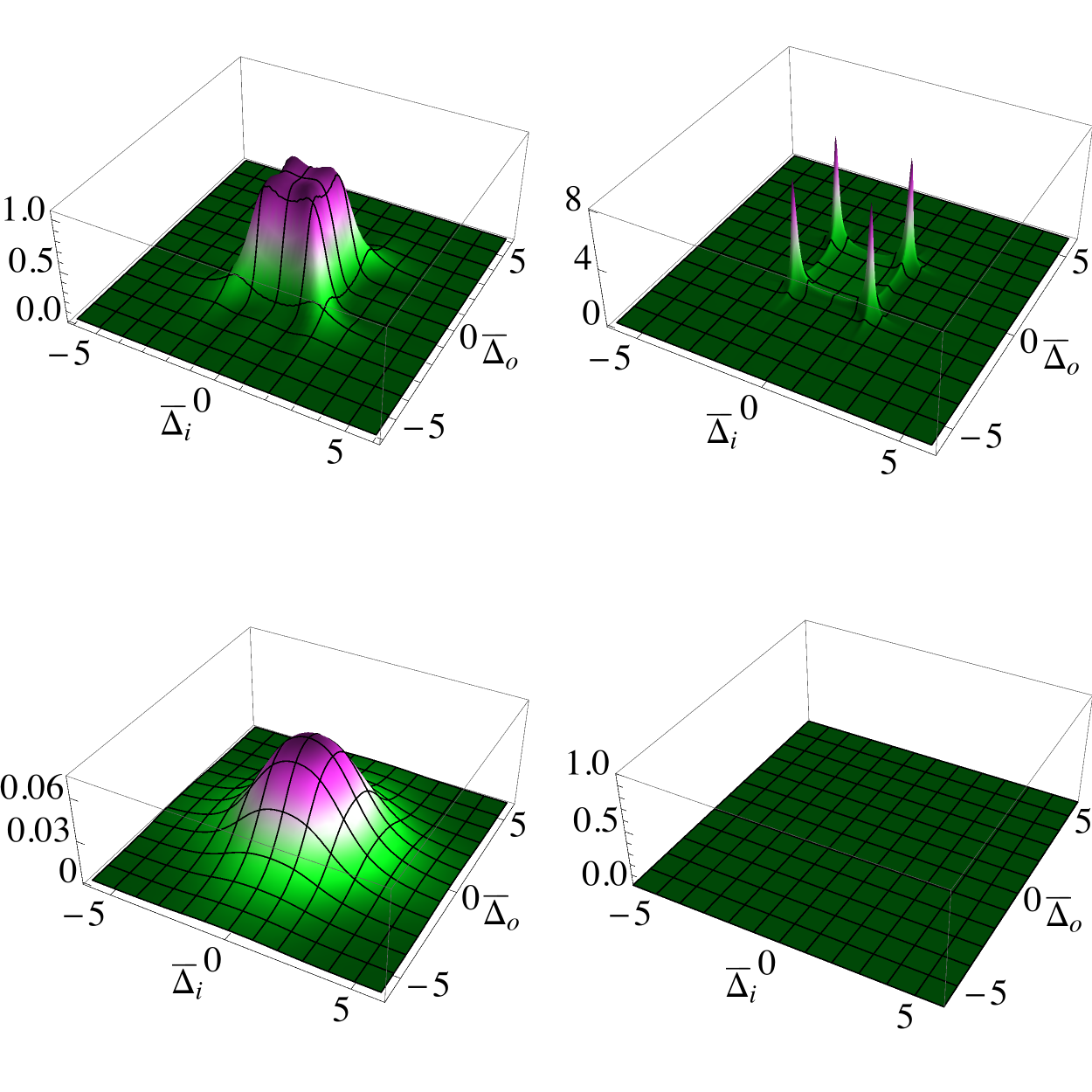}
\caption{(Color online) Plot of the normalized resonance fluorescence, $|B(k_{1,2}, p_{1,2})/\tau|^2$, assuming $\tau_1=\tau_2=\tau=1/\gamma$. $\Omega_c$, $\Omega_d$ and $\Delta_{i,o}$ are defined in the main text. Let $\bar{E}\equiv(E-2\Omega_c)\tau$ and  $\bar{\Delta}_{i,o}\equiv \Delta_{i,o} \tau$.  (a) $\bar{E}=3$, $\Omega_d=\gamma$. (b) $\bar{E}=3$, $\Omega_d=0.5\gamma$. (c) $\bar{E}=3$, $\Omega_d=0$. (d) $\bar{E}=0$.}
  \label{F:1}
\end{figure}
From the definition of $a_{\text{in}}(t)$ one can straightforwardly show that \cite{Fan2010}
\begin{align}
&\braopket{p^+_1}{\sigma^{(1,2)}_z(t) a _\text{in}(t)}{k_1 k^+_2}
=\frac{1}{\sqrt{2\pi}}\Bigg\{\frac{\me^{\mi(p_1-k_1-k_2)t}}{\pi} s^{*(1,2)}_{p_1}\left[s^{(1,2)}_{k_1}+s^{(1,2)}_{k_2}\right] \\
&-\me^{-\mi k_1t}\delta(k_2-p_1)-\me^{-\mi k_2t}\delta(k_1-p_1)\Bigg\}.
\end{align}
In solving for $\braopket{p^+_1}{\sigma^{(1)}_z(t)\sigma^{(2)}_-(t)}{k_1 k^{+}_2}$, we note
\begin{align}
&\braopket{p^+_1}{\sigma^{(1)}_z(t)\sigma^{(2)}_-(t)}{k_1 k^{+}_2}\\
&=2\bra{p^{+}_1}\sigma^{(1)}_+(t) \ket{0}\bra{0}\sigma^{(1)}_-(t) \sigma^{(2)}_-(t)\ket{k_1 k^{+}_2}-\bra{p^{+}_1}\sigma^{(2)}_-(t)\ket{k_1 k^{+}_2}. \label{E:2sigpsigm}
\end{align}
As a result, we are left with calculating the matrix element 
$\bra{0}\sigma^{(1)}_-(t) \sigma^{(2)}_-(t)\ket{k_1 k^{+}_2}$. From Eqs. \eqref{E:sig1b} and \eqref{E:sig2b}, we can derive an operator equation
\begin{align}
&\frac{\dif}{\dif t}\left[\sigma^{(1)}_-(t)\sigma^{(2)}_-(t)\right]=-\mi\left(\Omega_1+\Omega_2-\mi\frac{1}{\tau_1}-\mi\frac{1}{\tau_2}\right)\sigma^{(1)}_-(t)\sigma^{(2)}_-(t)\\&+\mi\sqrt{\frac{2}{\tau_1}}\sigma^{(1)}_z(t)\sigma^{(2)}_-(t)a _{\text{in}}(t)+\mi\sqrt{\frac{2}{\tau_2}}\sigma^{(2)}_z(t) \sigma^{(1)}_-(t)a _{\text{in}}(t), \label{E:sig1sig2}
\end{align}
where we have used the operator identity
\begin{align}
[a_\text{in}(t),\sigma^{(1,2)}_-(t)]=0\label{E:Commutator}.
\end{align}
[Equation \eqref{E:Commutator} is proved in Appendix \ref{AppA}.]  Equation \eqref{E:sig1sig2} can then be used to solve for $\bra{0}\sigma^{(1)}_-(t)\sigma^{(2)}_-(t)\ket{k_1 k^+_2}$.  At this stage, we have calculated all the prerequisite inhomogeneous terms in the coupled differential equations for  $\braopket{p^+_1}{\sigma^{(1,2)}_-(t)}{k_1 k^+_2}$. The resulting two-photon S-matrix is
\begin{align}
&\bra{0}a _{\text{out}}(p_1)a _{\text{out}}(p_2)a^{\dag} _{\text{in}}(k_2)a^{\dag} _{\text{in}}(k_1)\ket{0}\\
&=t_{p_1}t_{p_2}\left[\delta(k_1-p_1)\delta(k_2-p_2)+\delta(k_1-p_2)\delta(k_2-p_1)\right]\\
&+\Vast\{\frac{\mi}{\pi}\sqrt{\frac{2}{\tau_1}}s^{(1)}_{p_1}s^{(1)}_{p_2}\left[s^{(1)}_{k_1}+s^{(1)}_{k_2}\right]+\frac{\mi}{\pi}\sqrt{\frac{2}{\tau_2}}s^{(2)}_{p_1}s^{(2)}_{p_2}\left[s^{(2)}_{k_1}+s^{(2)}_{k_2}\right]\\
&+\left(\frac{s^{(1)}_{p_1}s^{(1)}_{p_2}+s^{(2)}_{p_1}s^{(2)}_{p_2}}{\pi\sqrt{\tau_1\tau_2}}\right) \frac{\sqrt{\frac{2}{\tau_1}}\left[s^{(2)}_{k_1}+s^{(2)}_{k_2}\right]+\sqrt{\frac{2}{\tau_2}}\left[s^{(1)}_{k_1}+s^{(1)}_{k_2}\right]}{\left(E_i-\Omega_1-\Omega_2+\mi\frac{1}{\tau_1}+\mi\frac{1}{\tau_2}\right)}\Vast\}\\
&\mspace{280mu}\times\delta(E_i-E_o) \label{E:sp1p2k1k2}
\end{align}
where $E_i\equiv k_1+k_2$ and $E_o \equiv p_1+p_2$.

Knowing the S-matrix of the chiral model which contains the photon-atom interaction, the full S-matrix for the waveguide system in Fig. \ref{figGeometry}, which has both the left and right going photons, can then be constructed straightforwardly \cite{Shen2007}.  In particular, the fluorescence spectra of the transmitted and reflected photons are described by the same last three terms in Eq. \eqref{E:sp1p2k1k2}.  Below, we will discuss the results for the waveguide system shown in Fig.\ref{figGeometry}.

Examining Eq. \eqref{E:sp1p2k1k2}, the first term describes an uncorrelated transport process where the energy of individual photons is conserved. The following three terms, which we collectively label $B(k_{1,2},p_{1,2})\delta(E_i-E_o)$, describe the fluorescence process where only the total energy of the photons, but not the individual energies, is conserved.  In particular, the second and third terms represent fluorescence from each individual atom.  The fourth term arises from the joint fluorescence in which both atoms are excited simultaneously, and contains a two-photon pole.  These different fluorescent pathways interfere coherently, leading to a complex set of interesting effects.

\pdfbookmark[5]{Results}{Results}
\section{Results}
\subsection{Fluorescence linewidth narrowing} 
In Fig. \ref{F:1} we plot the spectrum of fluorescence $|B(k_{1,2} , p_{1,2})/\tau|^2$, in the two-dimensional space spanned by the $\Delta_i\equiv (k_1-k_2)/2$ and $\Delta_o\equiv (p_1-p_2)/2$ axes. We assume the two atoms have identical waveguide coupling rates of $\gamma\equiv 1/\tau$, but generally different resonant frequencies, and define $\Omega_c\equiv\frac{\Omega_1+\Omega_2}{2}$; $\Omega_d\equiv\frac{\Omega_1-\Omega_2}{2}$. In Figs. \ref{F:1}(a)--\ref{F:1}(c), we plot the fluorescent spectrum for incident two photons with total energy $E_i=2\Omega_c+3\gamma$.  Examining Eq. \eqref{E:sp1p2k1k2}, we see that in the $\Delta_i\ -\ \Delta_o$ plane, the poles of $B(k_{1,2},p_{1,2})$ are the same as the poles of the atomic excitations $(s^{(1,2)}_{k_{1,2}}, s^{(1,2)}_{p_{1,2}})$ with single photon input.  The atomic excitation exhibits a sub-radiant state with poles corresponding to outgoing momenta $p_{1,2}\approx\Omega_c-\mi\Omega_d^2/\gamma$, and a super-radiant state with poles corresponding to outgoing momenta $p_{1,2}=\Omega_c-2\mi\gamma$.  Consequently when the atoms' detuning is smaller than their linewidth $(\Omega_d<\gamma)$, the fluorescent features are quite narrow [Fig. \ref{F:1}(b)].  At zero detuning, however, the sub-radiant state has zero linewidth and no longer couples to externally incident photons. As a result, the fluorescence is dominated by the super-radiant poles [Fig. \ref{F:1}(c)], and the fluorescent linewidth is doubled compared to a single atom's fluorescence. 
\pdfbookmark[1]{Fluorescence quenching and two-photon resonance}{Fluorescence quenching and two-photon resonance}
\subsection{Fluorescence quenching and two-photon resonance} 
Fluorescence is typically an unavoidable signature of interaction between an atom and multiple photons, since it arises from the inelastic scattering of one photon off an excited atom \cite{Shen2007}.  Here, as shown in Fig. \ref{F:1}(d), the fluorescence completely vanishes provided the total energy of the incident photons satisfies $E_i=\Omega_1+\Omega_2$.  Fluorescence quenching was previously noted in a driven three-level system \cite{Zhou1996}.  Here we show that quenching can occur in a system with two two-level systems.

The effect of fluorescence quenching is closely related to the existence of a two-photon pole in the S-matrix, which provides the necessary pathway to cancel the contribution from the fluorescence of individual atoms.  The existence of a two-photon pole indicates that the two atoms can be simultaneously excited by two photons, as long as the sum of the photon energy is near the sum of the transition energy of the two atoms.

Whether a two-atom system can exhibit a two-photon pole has been an interesting question. In Ref. \cite{Muthukrishnan2004} it is argued that in the absence of dipole-dipole interaction \cite{Varada1992}, simultaneous excitation is not possible classically.  The authors have also shown that simultaneous excitation is possible when the photon pair is frequency entangled, or when the two atoms interact via a quantized cavity field \cite{Kim1998}. Our contribution is in showing the connection between the two-photon pole in the joint excitation of the atoms and fluorescence quenching.
\begin{figure}
\includegraphics[scale=0.8]{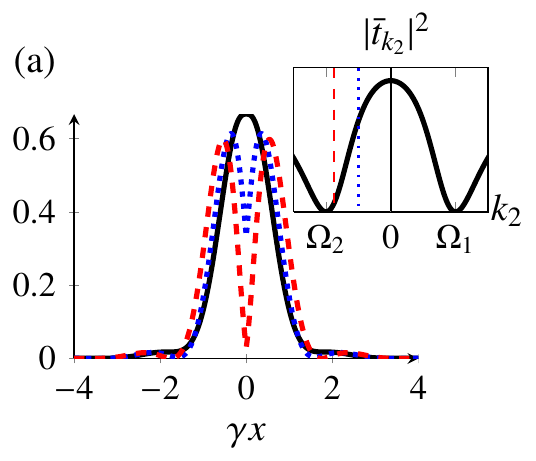}
\includegraphics[scale=0.8]{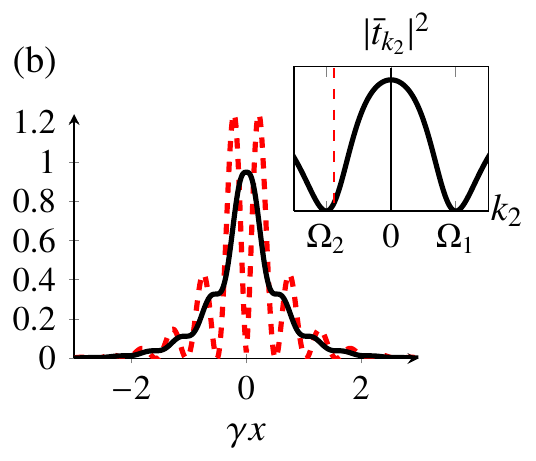}
\includegraphics[scale=0.8]{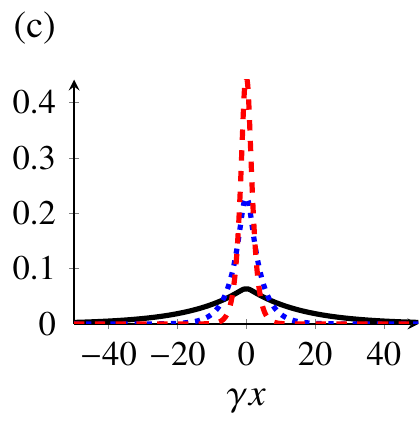}
\caption{(Color online) $P^{(R)}_2(x)$, the joint detection probability density of the transmitted two-photon state. (a) $k_1=\Omega_1$,  $\Omega_d=2\gamma$, $(k_2-\Omega_2)=\{2, 1, \frac{1}{4}\}\gamma$ in solid black, dotted blue and dashed red curves, respectively. (b) $k_1=\Omega_1$, $\Omega_d=6\gamma$, $(k_2-\Omega_2)=\{6,0.25\}\gamma$ in solid black and dashed red curves, respectively.  In (a) and (b), insets show the frequency $k_2$ overlaid on the single-photon transmission $|\bar{t}_k|^2$.   (c) $k_1=\Omega_1$, $k_2=\Omega_c = \{0.75,0.5,0.25\}\gamma $ for the black (solid), blue (dotted) and red (dashed) curves, respectively.} \label{F:2}
\end{figure}
\subsection{Generation and control of two-photon bound states}
For an incoming state comprising two right-going photons with individual energies $k_1$ and $k_2$, the resultant transmitted two-photon state is:
\begin{align}  
&\ket{\psi_{R}}=\int \dif x_1\dif x_2 \left[\bar{t}_{k_1}\bar{t}_{k_2}S_{k_1,k_2}(x_1,x_2)+\textstyle{\frac{1}{4}}H(x_1,x_2)\right]
\ket{ x_1,x_2}_{RR}\label{E:psioutRR}.
\shortintertext{Here $\bar{t}_k=(t_k+1)/2$ is the single-photon transmission amplitude. $S_{k_1,k_2}(x_1,x_2)=\frac{1}{2\pi\sqrt{2}}\left[\me^{\mi(k_1x_1+k_2x_2)}+\me^{\mi(k_1x_2+k_2x_1)}\right]$, and $\ket{x_1,x_2}_{RR}=\frac{1}{\sqrt{2}}c^{\dag}_{R}(x_1)c^{\dag}_{R}(x_2)\ket{0}$ where $\left[c_R(x),c^{\dag}_R(x')\right]=\delta(x-x')$.} 
&H(x_1,x_2)\equiv\\
&\phantom{+} \sqrt{\frac{2}{\tau_1}}\frac{\soko+\sokt}{\pi}\left[F_{1}(x_1,x_2)\left(1-\frac{\mi}{D_a \tau_2}\right)+F_{2}(x_1,x_2)\left(\frac{-\mi}{D_a \tau_2}\right)\right]\\
&+\sqrt{\frac{2}{\tau_2}}\frac{\stko+\stkt}{\pi}\left[F_{1}(x_1,x_2)\left(\frac{-\mi}{D_a \tau_1}\right)+F_{2}(x_1,x_2)\left(1-\frac{\mi}{D_a \tau_1}\right)\right]\label{E:H}
\end{align}
where
\begin{align}
&F_{1,2}(x_1,x_2)\equiv\frac{\sqrt{2}\me^{\mi E_i\left(\frac{x_1+x_2}{2}\right)}}{\tau_{1,2}D_a D_b} \times\\
&\left\{ \frac{\me^{\mi D_1 \abs{x_1-x_2}}\left[D_1^2 - (\frac{E_i}{2}  - \Omega_{2,1})^2\right]}{D_a + D_b} 
-\frac{\me^{\mi D_2 \abs{x_1-x_2}}\left[D_2^2 - (\frac{E_i}{2}  - \Omega_{2,1})^2\right]}{D_a - D_b}\right\} 
\end{align}
with $D_{1,2}\equiv \left(D_{a} \pm D_{b}\right)/2$, $D_a \equiv E_i-2\Omega_c+\mi/\tau_1+\mi/\tau_2$,
$D_b \equiv [4\Omega_d^2+4\mi\Omega_d(1/\tau_1-1/\tau_2)-(1/\tau_1+1/\tau_2)^2]^{1/2}$.  $\ket{\psi_{R}}$ contains an uncorrelated-transport extended plane-wave term [ $\bar{t}_{k_1}\bar{t}_{k_2}S_{k_1,k_2}(x_1,x_2)$ ], and a bound-state term, [$\frac{1}{4}H(x_1,x_2)$], which arises directly from fluorescence.

Since $|_{RR}\bra{x_1,x_2}\psi_R\rangle|^2=\bra{\psi_R}a^{\dag}_{R}(x_2)a^{\dag}_{R}(x_1)a_{R}(x_2)a_{R}(x_1)\ket{\psi_R}$ is proportional to the joint-detection probability density $P^{(R)}_2(x_1-x_2)$, this wavefunction therefore can be experimentally probed in a Hanbury Brown-Twiss coincidence measurement, with $x=x_1-x_2$ being the difference in optical path length from each detector to the beam splitter.  

The transmission amplitude $\bar{t}_k$ vanishes when $k=\Omega_{1,2}$, making it possible to eliminate the uncorrelated part of $\ket{\psi_{R}}$ by choosing $k_1=\Omega_1$, which we do in Fig. \ref{F:2}.  By doing so, the transmitted two-photon wavefunction is entirely described by the function $H(x_1,x_2)$, which represents a two-photon bound state that decays with respect to the photon spacing $x=x_1-x_2$. 

When $k_2=\Omega_c$, $P^{(R)}_2(x)$ exhibits a bunching behavior with a global maximum at $x=0$, as shown in Fig. \ref{F:2}(a).  As one decreases $k_2$ from this value, a local minimum at $x=0$ starts to develop, indicating anti-bunching \cite{Mandel1995}, which increases as $k_2\to\Omega_2^+$.  In this system, therefore, the two-photon bound-state can exhibit either a bunching or an anti-bunching behavior.  This is in contrast to the one-atom case \cite{Shen2007}, where the bound state by itself is always bunched.  It is also different from the typical resonance fluorescence experiments with classical input state, where anti-bunching is observed \cite{Kimble1977}.

In Fig. \ref{F:2}(b) the transmitted bound state is plotted for a large atomic detuning of $\Omega_d=6\gamma$, exhibiting spatial oscillation or quantum beats with a period of $2\pi/|D_b|$.  A two-photon plane wave $S_{k_1,k_2}(x_1,x_2)$ has a $P^{(R)}_2(x_1-x_2)$ that oscillates with a spatial period of $k_1-k_2$ where $k_{1,2}$ are the energy of the two individual photons.  Similarly, the quantum beats here indicate that the bound state is two colored, with the single-photon energies centered approximately at $\Omega_1$ and $\Omega_2$. Previously, the second-order correlation function of two atoms driven by a coherent laser field was shown to display similar oscillations which are governed by the atomic detuning \cite{Ficek2002}.  Here we note the connection between these oscillations and the spatial wavefunction of the two-photon bound state.

  Finally, we note that the spatial extent of the bound state is strongly dependent upon the atomic detuning $\Omega_d$, in the region where $\Omega_d<\gamma$.  In Fig. \ref{F:2}(c), $P^{(R)}_2(x)$ is plotted for three values of $\Omega_d$ satisfying $\Omega_d/\gamma<1$, with $k_2-\Omega_2=\Omega_d$.  For these values of $\Omega_d$, the sub-radiant poles of the scattering matrix, whose imaginary parts strongly depend on $\Omega_d$, dominate the bound-state response.  Consequently, as shown in Fig. \ref{F:2}(c), the spatial extent of the bound state can be significantly wider than the one-atom bound state.  Taken together, Figs. \ref{F:2}(a)--(c) demonstrate the significant ability of the two-atom system to control the properties of the two-photon bound state, including its spatial extent, quantum beats and statistics.  
  
\subsection{Experimental considerations}
We end by discussing some practical considerations relevant in experimental study of this system. All predicted effects in the paper require that the atomic resonance frequencies be close to each other.  With respect to the tuning of atom resonance frequencies, we note that tuning of quantum dots in the optical frequency range has been achieved via magnetic fields \cite{Haft2002}, the application of a dc voltage \cite{Hogele2004}, and through thermal heating \cite{Faraon2007}.  In the microwave, the ability to independently tune the transition frequencies of individual qubits has recently been demonstrated \cite{Gambetta2011}. It is conceivable that some of these techniques can be further developed to achieve independent tuning of two closely spaced qubits, particularly in the microwave frequency range.

We also assumed that each atom predominantly couples to the waveguide, i.e. that the system has a high $\beta$ factor.  (The $\beta$ factor measures the fraction of the spontaneous emission going into the guided mode.)  A high $\beta$ factor has been reported in the experiments of Ref. \cite{Lund-Hansen2008,Thyrrestrup2010}.  A value of $\beta<1$ may be accounted for by replacing each resonant frequency $\Omega_{1,2}$ with $\Omega_{1,2}-i\gamma^{(ng)}_{1,2}$ where $\gamma^{(ng)}_{1,2}$ are the respective coupling rates of each atom into non-guided modes.  Consequently, a non-unity $\beta$ will lead to the further narrowing of the fluorescence features in Fig. \ref{F:1}(b), and to the further broadening of the fluorescent features in Fig. \ref{F:1} (a),(c). 

The main result of this paper [i.e., the two-photon S-matrix in Eq. \eqref{E:sp1p2k1k2}], is valid for two atoms with either identical or non-identical atom-waveguide coupling rates.  As for the predicted effects, linewidth narrowing and the various properties of the bound state persist for atoms with non-identical waveguide coupling rates. Complete fluorescence quenching, however, requires that the coupling rates be identical.  For this purpose, we note that, in the microwave frequency range, Ref. \cite{Gambetta2011} has demonstrated the capability of tuning the qubit coupling rate.

We have calculated the response of the system to a two-photon Fock state input.  The predictions about the properties of the transmitted two-photon state can be observed by a correlation measurement with a weak coherent state input, which can be generated with an attenuated laser beam \cite{Shen2007}.  The frequency linewidth of such a beam can in principle be made narrower than any of the spectral features that we predict here. Alternatively, we note the recent development of deterministic single-photon sources \cite{Kuhn2002} as well as the demonstration of single photon pulses with arbitrary temporal shapes \cite{Kolchin2008}, both of which may facilitate the experimental study of this system.

In our calculations, we have presented the S-matrix for two incident photons, each having well-defined energy $k_1$ and $k_2$.  Any experiment, of course, uses a source with non-zero spectral bandwidth.  The spectrum of the output state is a product of the scattering matrix and the spectrum of the input state, and thus can be directly calculated using Eq. \eqref{E:sp1p2k1k2}.  Since the fluorescence is already in the spectral domain, the use of a non-zero bandwidth input should not affect the results of such spectral measurement.  In Fig. \ref{F:2}, we have shown that a pure two-photon bound state can be generated when one of the incident photons is on resonance with one of the atoms.  The use of a photon pulse would therefore result in a background amplitude due to uncorrelated transport, in addition to the bound state.  This background, however, can be made to be very weak, provided that the incident spectrum is significantly narrower than the atomic linewidth. 
\appendix
\section{PROOF OF EQ. (\ref{E:Commutator}) }
\label{AppA}
From the Hamiltonian in Eq. \eqref{E:Htotal}, an equation of motion for $a_k(t)$ may be derived
\begin{align}
\frac{\dif}{\dif t} a_k(t)=-\mi a_k(t)-\mi V_1\sigma^{(1)}_-(t)-\mi V_2\sigma^{(2)}_-(t)\label{E:akdot}
\end{align}
Equation \eqref{E:akdot} may be solved by integrating from time $t_0$:
\begin{align}
a_k(t)=a_k(t_0)\me^{-\mi k(t-t_0)}-\mi V_1\int_{t_0}^{t}\dif t'\me^{\mi k(t'-t)}\sigma^{(1)}_-(t')-\mi V_2\int_{t_0}^{t}\dif t'\me^{\mi k(t'-t_0)}\sigma^{(2)}_-(t')\label{E:aktApp}
\end{align}
Multiplying Eq. \eqref{E:aktApp} by $\frac{1}{\sqrt{2\pi}}$ and integrating over $k$ while taking the limit $t_0\to -\infty$, we obtain:
\begin{align}
a_{\text{in}}(t)=\frac{1}{\sqrt{2\pi}}\int \dif k \ a_k(t)+\mi \sqrt{\frac{1}{2\tau_1}}\sigma^{(1)}_-(t)+\mi \sqrt{\frac{1}{2\tau_2}}\sigma^{(2)}_-(t)\label{E:PhiApp}
\end{align}
where we have used the definition of $a_{\text{in}}(t)$ from Sec. \ref{System hamiltonian and equations of motion}.  It then follows that
\begin{align}
\left[a_{\text{in}}(t), \sigma^{(1)}_-(t)\right]=\frac{1}{\sqrt{2\pi}}\int \dif k \left[a_k(t),\sigma^{(1,2)}_-(t)\right]+\mi \sqrt{\frac{1}{2\tau_1}}\left[\sigma^{(1)}_-(t),\sigma^{(1,2)}_-(t)\right]+\mi \sqrt{\frac{1}{2\tau_2}}\left[\sigma^{(2)}_-(t),\sigma^{(1,2)}_-(t)\right]=0 \label{E:Phi2App}
\end{align}

\bibliography{TwoAtoms}

\end{document}